\renewcommand{\theequation}{\thesection.\arabic{equation}}

\documentstyle[12pt]{article}    
\renewcommand{\theequation}{\thesection.\arabic{equation}}

\begin{document}

\Large
\centerline{\bf $Q^2$ evolution of chiral-odd twist-3 distribution
$e(x,Q^2)$}
\normalsize
\setlength{\baselineskip}{0.3in}

\vspace{1.5cm}
\centerline{Yuji Koike and N. Nishiyama}

\centerline{\it Graduate School of Science and Technology, 
Niigata University,
Ikarashi, Niigata 950-21, Japan}

\vspace{3cm}

\centerline{\bf Abstract}

We study the $Q^2$ dependence of the 
chiral-odd
twist-3 distribution $e(x,Q^2)$.  
The anomalous dimension matrix for the 
corresponding twist-3 operators
is calculated in the one-loop level.  This study completes 
the calculation
of the anomalous 
dimension matrices for all the twist-3 distributions
together with the known results for the other twist-3
distributions $g_2(x,Q^2)$ and $h_L(x,Q^2)$.  We also have confirmed
that in the large $N_c$ limit the $Q^2$-evolution of $e(x,Q^2)$ is
wholely governed 
by the lowest eigenvalue of the anomalous dimension matrix
which takes a very simple analytic form as in the case of
$g_2$ and $h_L$.

\vspace{1cm}

\newpage
\section{Introduction}
\setcounter{equation}{0}
\renewcommand{\theequation}{\arabic{section}.\arabic{equation}}

Ongoing plans of
high energy collider experiments are going on their way 
toward more precision 
measurements and more variety of spin-dependent 
observables in many (semi) inclusive processes.
Correspondingly, increasing 
attention is paid on the power corrections due to
the higher twist-effects which represent 
parton correlations in the target.
Recent measurement of the nucleon's $g_2$ 
structure function by the E143 
collaboration\,\cite{E143} anticipates a forthcoming significant 
progress of the twist-3 physics.  Under this circumstance,
QCD study on the scale dependence of 
various distribution and fragmentation 
functions is of great interest.

The nucleon has three twist-3 
distributions $g_2$, $h_L$ and $e$\,\cite{JJ}.  
$g_2$ is chiral-even and 
the other two are chiral-odd.   $e$ is spin-independent 
and the other two are spin-dependent. 
Compared to $e(x,Q^2)$, $g_2$ and $h_L$ have 
more chance to be measured experimentally 
since they become leading contribution
to proper asymmetries in the polarized 
deep inelastic scattering (DIS) and Drell-Yan processes,
respectively.  
Their $Q^2$ evolution has been studied in \cite{BKL,ABH,KUY} 
for $g_2$ and in \cite{KT,BBKT} for $h_L$.

Similarly to the distribution functions, there are three twist-3 
fragmentation functions which describes hadronization 
processes of partons 
in semi-inclusive processes\,\cite{JJ2}, $\hat{g}_2(z,Q^2)$, 
$\hat{h}_L(z,Q^2)$
and $\hat{e}(z,Q^2)$. (Their naming is parallel to the corresponding 
distribution functions.)  In the inclusive  
pion production in the transversely polarized DIS,  
the chiral-odd fragmentation function $\hat{e}$ of the pion
appears as a leading contribution together with $h_1$ (twist-2)
of the nucleon. 
Although the $Q^2$ 
evolution of the twist-2 fragmentation functions is 
known to be obtained from
that of 
the corresponding distributions in the one-loop level (Gribov-Lipatov 
reciprocity\,\cite{GL}), no such 
relation is known for the higher twist 
fragmentation functions.

In this paper we investigate the $Q^2$ evolution of $e(x,Q^2)$.  
Theoretically, this completes the calculation of the 
anomalous dimension 
matrices of all the 
twist-3 distributions.  Phenomenologically, we expect 
it will shed light on the $Q^2$ evolution of $\hat{e}(z,Q^2)$, 
anticipating the day when their relation is clarified.

The outline of this paper is the following:
In section 2, we briefly recall 
the twist-3 operators for $e(x,Q^2)$\,\cite{JJ}.
Similarity to the $h_L$ case is reminded.  In section 3, we present
the calculation of the one-loop anomalous dimension
matrix for $e(x,Q^2)$.  The method follows that of \cite{KT}. 
In section 4, the $Q^2$ evolution
of $e(x,Q^2)$ in the large $N_c$ limit is
discussed.
This part is a recapitulation of \cite{BBKT} in our language.
Section 5 is devoted to a brief summary of our results.
Appendix contains the contributions from each 
one-loop Feynman diagram.

\section{Twist-3 operators for $e(x,Q^2)$}
\setcounter{equation}{0}
\renewcommand{\theequation}{\arabic{section}.\arabic{equation}}

The chiral-odd twist-3 distribution function 
$e(x,Q^2)$ 
is defined by the relation\,\cite{JJ}
\begin{equation}
\int { d\lambda  \over 2 \pi } e^{i\lambda x} \langle P |
\bar{\psi}(0)  
\psi(\lambda n)|_Q |P \rangle 
= 2 Me(x,Q^2),
\label{eq2.1}
\end{equation}
where $|P\rangle$ is the nucleon (mass $M$)
state with momentum $P$. 
Two light-like vectors $p$ and $n$ defined by the relation,
$P=p+{M^2\over 2}n$, $p^2=n^2=0$, $p\cdot n=1$,
specify the Lorentz frame of the system.
Gauge-link operators are implicit in (\ref{eq2.1}).
Taking the moments of (\ref{eq2.1}) with respect to $x$, one 
can express the moments of $e(x,Q^2)$ in terms of
the nucleon matrix elements of the twist-3 operators
$V^{\mu_1\cdots\mu_n}$:
\begin{eqnarray}
{\cal M}_n\left[e(Q^2)\right] &=& e_n(Q^2),
\label{eq2.2}\\
\langle PS | V^{\mu_1 \mu_2 \cdots \mu_n} | PS \rangle &=& 2 e_n M 
(P^{\mu_1}P^{\mu_2}\cdots P^{\mu_n} -{\rm traces}),
\label{eq2.3}\\
V^{\mu_1 \mu_2 \cdots \mu_n} &=&
 {\cal S}_n \bar{\psi}iD^{\mu_1}iD^{\mu_2} \cdots
iD^{\mu_n}\psi - {\rm traces},
\label{eq2.4}
\end{eqnarray}
where 
$ {\cal M}_n\left[ e(Q^2)\right]\equiv 
\int_{-1}^1\,dx\,x^n e(x,Q^2)$ and
the covariant derivative 
$D_\mu = \partial_{\mu} -igA_\mu$ restores
the gauge invariance.
As in (\ref{eq2.3}) and (\ref{eq2.4}), we often suppress
the explicit scale dependence. 
Following a common wisdom we introduce
a null vector $\Delta_\mu$ ($\Delta^2=0$) to kill 
the trace terms of
$V^{\mu_1\cdots\mu_n}$ and introduce the notation
$V_n\cdot\Delta\equiv V^{\mu_1\cdots\mu_n}
\Delta_{\mu_1}\cdots\Delta_{\mu_n}$
for convenience, and similarly for other operators
in the following.
Using the 
relation $D^{\mu}=\frac{1}{2} \left\{
\rlap/{\mkern-1mu D}, \gamma^{\mu} \right\}$ and
$[D_\mu, D_\nu] = -igG_{\mu\nu}$ with 
the gluon field strength $G_{\mu \nu}$,
$V_n \cdot \Delta$ can be recast into the following form:
\begin{equation}
V_n\cdot \Delta = \sum^{n}_{l=2} U_{n,l}\cdot \Delta
+ N_n \cdot \Delta + E_n \cdot \Delta,
\label{eq2.5}
\end{equation}
where 
\begin{eqnarray}
U_l^{\mu_1 \cdot\cdot\cdot \mu_n} 
&=& {1 \over 2}
{\cal S}_n \bar{\psi}\sigma^{\alpha\mu_1} 
 iD^{\mu_2}\cdots gG^{\mu_l}_{\ \, \alpha}
\cdots iD^{\mu_n} \psi - {\rm traces},
\label{eq2.6}\\
N^{\mu_1 \cdots \mu_n}&=&{\cal S}_n m_q \bar{\psi} \gamma^{\mu_1}
iD^{\mu_2} \cdots iD^{\mu_n} \psi - {\rm traces},
\label{eq2.7}\\
E^{\mu_1
\cdot\cdot\cdot \mu_n} 
&=& {1 \over 2} {\cal S}_{n}
\left[ \bar{\psi}(i\rlap/{\mkern-1mu D} -m_q ) 
\gamma^{\mu_{1}}
iD^{\mu_{2}} \cdots iD^{\mu_{n}} \psi +
\bar{\psi} \gamma^{\mu_{1}}
iD^{\mu_{2}} \cdots iD^{\mu_{n}}(i\rlap/{\mkern-1mu D} 
-m_q)\psi \right]
- {\rm traces}.\nonumber\\
\label{eq2.8}
\end{eqnarray}
$U_l$ contains $G_{\mu\nu}$ explicitly,
which indicates that $e(x)$ represents the quark-gluon 
correlations in the
nucleon.  $N$ is the quark mass times the twist-2
operator which contributes to the moments of $f_1$ 
structure function
familiar in the spin-averaged DIS.
$E$ is the EOM (equation of motion) operator
which vanishes by use of the QCD equation-of-motion.
Although the physical matrix elements of EOM operators
vanish, one needs to take into account the mixing
with $E$ to carry out the renormalization of 
$U$ and $N$,
as is discussed in \cite{KUY,KT} in the context of
the renormalization of $g_2$ and $h_L$.
From (\ref{eq2.5}) one sees that $U_l$ appears in the form
of
\begin{eqnarray}
R_{n,l}^{\mu_1 \cdot\cdot\cdot \mu_n} 
= U_{n-l+2}^{\mu_1 \cdot\cdot\cdot \mu_n}
+ U_l^{\mu_1 \cdot\cdot\cdot \mu_n}, \ \ \ \left(
l=2,...,\left[{n \over 2}\right]+1 \right).
\label{eq2.9} 
\end{eqnarray}
By this combination, $R_{n,l}$ has a definite charge 
conjugation $(-1)^n$.
Readers may recall the similarity between the present
\{$U_l$, $R_{n,l}$\} and \{$\theta_l$, $R_{n,l}$\} which appeared
in $h_L$\,\cite{JJ,KT}.  
In fact presence of 
$\gamma_5$ in $\theta_l$ is the mere difference from $U_l$.
$R_{n,l}$ of $h_L$ is defined as
$\theta_{n-l+2}-\theta_l$ and has a charge conjugation
$(-1)^{n+1}$ which is opposite to the above $R_{n,l}$ 
in (\ref{eq2.9}).
We will see the similarity in the renormalization constants 
between $h_L$ and $e$ in the following.

\section{$Q^2$ evolution of $e(x,Q^2)$}
\setcounter{equation}{0}
\renewcommand{\theequation}{\arabic{section}.\arabic{equation}}

For the renormalization of $e(x)$, we closely follow the method of
\cite{KT} which discussed the renormalization of $h_L$.  So we
omit the detail in the following.
Owing to the chiral-odd nature, $e(x)$ does not mix with
the gluon-distribution.
We choose
$R_{n,l}$ ($l=2,\ldots, \left[ \frac{n}{2} \right]+1$), $E$, $N$
as a basis of the operators.  For the renormalization
of $R_{n,l}$, we calculate the one-loop
correction to the three-point
function $F_\mu (p,q,p-q)$ defined by
\begin{eqnarray}
 & & F_\mu(p,q,k)(2\pi)^4 \delta^4(p+k-q) G(p)G(q) D(k) \nonumber\\
 & & \ \ \ \  = \int d^4x d^4y d^4z e^{ipx} e^{-iqy} e^{ikz} 
 \langle T \{ {\cal O} \psi(x) \bar{\psi}(y) A_\mu(z) \}
\rangle,\ \ \ \ ({\cal O}=R_{n,l},\ N,\ E)
\label{eq3.1}
\end{eqnarray}
where $G$ and $D$ are, respectively, the quark and gluon 
propagators. (We suppressed the Lorentz and spinor indices
for simplicity.)  
In order to take into account the mixing with the EOM
operator properly, 
we use off-shell kinematics for the external lines.
The calculation is 
done with the Feynman gauge for the gluon propagator 
(of course the result should be independent of the gauge) and  
the MS scheme is adopted with the dimensional regularization.
The three-point basic vertices shown in Fig. 1(a) for 
$R_{n,l}\cdot\Delta$, $E_n\cdot\Delta$, $N_n\cdot\Delta$
are calculated to be
\begin{eqnarray}
{\cal R}^{(3)}_{n, l, \mu} &=&
-i{g \over 2} \sigma^{\alpha\lambda}\Delta_\lambda 
({\hat p}^{n-l}{\hat q}^{l-2}+{\hat p}^{l-2}{\hat q}^{n-l})
( -{\hat k} g_{\alpha\mu} 
+ k_\alpha \Delta_\mu ) t^a,
\label{eq3.2}\\
{\cal E}^{(3)}_{n, \mu} 
& =&{g \over 2} \left[ \gamma_\mu \rlap/{\mkern-1mu
\Delta} 
{\hat q}^{n-1} 
+ \rlap/{\mkern-1mu \Delta} {\hat p}^{n-1} 
\gamma_\mu 
+ \Delta_\mu \sum_{j=2}^n
(\rlap/{\mkern-1mu p} -m_q)  \rlap/{\mkern-1mu
\Delta} 
{\hat p}^{j-2} {\hat q}^{n-j}
\right.\nonumber\\
& & \left. 
\ \ \ \ \ \ + \Delta_\mu \sum_{j=2}^n 
\rlap/{\mkern-1mu \Delta} (\rlap/{\mkern-1mu q} -m_q)
{\hat p}^{j-2}  {\hat q}^{n-j}
\right]t^a,
\label{eq3.3}\\
{\cal N}^{(3)}_{n, \mu} 
& =& m_q g\Delta_{\mu} \rlap/{\mkern-1mu \Delta}
\sum_{j = 2}^{n} 
{\hat p}^{j-2}{\hat q}^{n-j} t^{a},
\label{eq3.4}
\end{eqnarray}
where $k=q-p$, ${\hat p}=p\cdot\Delta$
for an arbitrary four vector $p$ and $t^a$ is 
the color matrix normalized as ${\rm Tr}(t^a t^b)= \frac{1}{2}
\delta^{ab}$.
One-loop diagrams for the above $F_\mu$ are
shown in Fig. 2. In calculating these diagrams, we need a 
Feynman rule for 
the four-point basic vertex of $R_{n,l}\cdot \Delta$ 
shown in Fig. 1(b).
It is given by
\begin{eqnarray}
& & {g^2 \over 2} \left[ f^{abc}t^c \sigma^{\alpha\lambda}
\Delta_\lambda  
\Delta_\mu g_{\alpha\nu} {\hat p}^{n-l} {\hat q}^{l-2}
\right. \nonumber\\[8pt]
& & \left.
- i\sigma^{\alpha\lambda}
\Delta_\lambda  
t^at^b \sum_{j=2}^{n-l+1} 
{\hat p}^{j-2} \Delta_\mu ({\hat p} + {\hat k})^{n-l+1-j}
(-{\hat k'}g_{\alpha\nu} + k'_\alpha \Delta_\nu ) {\hat q}^{l-2}
\right. \nonumber\\[8pt]
& & \left. 
- i\sigma^{\alpha\lambda}
\Delta_\lambda 
t^at^b \sum_{j=n-l+3}^n
{\hat p}^{n-l} \Delta_\nu ({\hat p} + {\hat k})^{j-n+l-3}
(-{\hat k}g_{\alpha\mu} + k_\alpha \Delta_\mu ) {\hat q}^{n-j}
\right. \nonumber\\[8pt]
& & \left. + (\mu \leftrightarrow \nu, k \leftrightarrow k', 
a \leftrightarrow b)
\right]  + ( l \rightarrow n-l+2).
\label{eq3.5}
\end{eqnarray}
For the actual calculation of the one-loop diagrams, we 
introduce
a vector $\Omega_\mu$
with the condition $\Omega\cdot\Delta=0$, and 
consider $F_\mu\Omega^\mu$\,\cite{KT}.
This way mixing with the gauge noninvariant
EOM operators can be avoided and the
calculation is greatly simplified.  (See \cite{KT} for the detail.) 
The contraction of ${\cal R}^{(3)}_{n,l,\mu}$ and
${\cal E}^{(3)}_{n,l,\mu}$ with $\Omega_\mu$ leads to
\begin{eqnarray}
{\cal R}_{n,l}^{(3)}\cdot \Omega 
&=& {ig \over 2} \Omega_\alpha \sigma^{\alpha\lambda}
\Delta_\lambda  ({\hat q}-{\hat p}) \left(
{\hat p}^{n-l} {\hat q}^{l-2} + {\hat p}^{l-2}{\hat q}^{n-l} 
\right)t^a,
\label{eq3.6}\\
{\cal E}_n^{(3)}\cdot\Omega
&=&-{ig \over 2} \Omega_\alpha \sigma^{\alpha\lambda}
\Delta_\lambda  \left(
{\hat q}^{n-1} - {\hat p}^{n-1} \right) t^a.
\label{eq3.7}
\end{eqnarray}
The mixing coefficients between $R_{n,l}$ and \{$N$, $E$\}, $Z_{lE}$
and $Z_{lN}$, can also be obtained from
the one-loop correction to the two point function 
shown in Fig. 3 (b), (c),
giving a consistency check.
The two point basic vertices
for $E$ and $N$ are
\begin{eqnarray}
{\cal E}_n^{(2)}&=& {1 \over 2} {\hat p}^{n-1}
\left( \rlap/{\mkern-1mu \Delta} \rlap/{\mkern-1mu p} 
+ \rlap/{\mkern-1mu p} \rlap/{\mkern-1mu \Delta} 
- 2m_q \rlap/{\mkern-1mu \Delta} \right),
\label{eq3.8}\\
{\cal N}_n^{(2)}&=& m_q {\hat p}^{n-1}
\rlap/{\mkern-1mu \Delta}  
\label{eq3.9}.
\end{eqnarray}

Using these Feynman rules for the vertices, 
one can calculate the one-loop diagrams shown in Figs. 2 and 3.
Actual calculation is very tedious and will be described 
in the Appendix
in detail.  Taking into account the wave function
renormalization for the fields and the renormalization 
for $m_q$ and $g$
which explicitly appear in the vertices\,\cite{KT},
we eventually obtained 
the renormalization constants $Z_{ij}$ 
among $R_{n,l}$, $N$ and $E$
in the following matrix form:
 \begin{eqnarray}
\left(\matrix{R_{n,l}^B\cr
              E_n^B\cr
              N_n^B\cr}\right)=
\left(\matrix{Z_{lm}(\mu)&Z_{lE}(\mu)&Z_{lN}(\mu)\cr
              0&Z_{EE}(\mu)&0\cr
              0&0&Z_{NN}(\mu)\cr}\right)
\left(\matrix{R_{n,m}(\mu)\cr
              E_n(\mu)\cr
              N_n(\mu)\cr}\right),
\ \ \ \left(l,m = 2,\cdot\cdot\cdot,\left[{n \over 2}
\right]+1\right).
\label{eq3.10}
\end{eqnarray}
$Z_{ij}$ can be expressed as
\begin{eqnarray}
Z_{ij} = \delta_{ij}+ {g^2 \over 16\pi^2 \varepsilon} Y_{ij}
\ \ \ \ \left(i,j= 2,\cdot\cdot\cdot,\left[{n \over 2}
\right]+1 ,
E, N\right)
\label{eq3.11}
\end{eqnarray}
with $\varepsilon={4-d \over 2}$ 
($d$ is the space-time dimension in the dimensional regularization) 
and the constants $Y_{ij}$ given below.  To write down $Y_{ij}$
we define a symbol $\langle {n\over 2} \rangle$
as $\langle {n\over 2}\rangle\equiv [{n\over 2}]={n\over 2}$
for an even $n$ and
$\langle {n\over 2}\rangle\equiv [{n\over 2}]+1={n+1\over 2}$
for an odd $n$.  Then $Y_{ij}$ is given as follows:
\begin{eqnarray}
Y_{lm} &=&  C_G 
\left[
   { (m-1)(m+l) \over 2(l-m)[l-1]_2 } 
  +{ (m-1)(m+n-l+2) \over 2(n-l-m+2)[n-l+1]_2 }
  -{ 1 \over l } 
  -{ 1 \over n-l+2 } \right. \nonumber\\[8pt]
& &\left. + { (l-3)(l+1-m) \over  2[l-1]_3 }
  +{ l+2 \over 2[l-1]_2 }
  +{ (n-l-1)(n-l-m+3) \over 2[n-l+1]_3 }
  +{ n-l+4 \over 2[n-l+1]_2 }
\right]  \nonumber\\[8pt]
& &+ (2C_F -C_G) \left[ 
{ (-1)^{n-l-m}\ _{n-l}C_{m-2} \over 
(n-l+2-m)\ _{n-m+1}C_{l-1} }  
+ { (-1)^{l-m}\ _{l-2}C_{l-m} \over 
(l-m)\ _{n-m+1}C_{l-m} }
\right. \nonumber\\[8pt]
& & \left. - 2 (-1)^m \left( { _{l-1}C_{m-1} \over [l-1]_3 }
+ { _{n-l+1}C_{m-1} \over [n-l+1]_3 } \right)  \right]\nonumber\\
& &(2 \le l \le \left[ { n \over 2 } \right]+1,
\ \ 2 \le m \le l-1),
\label{eq3.12}
\end{eqnarray}
\begin{eqnarray}
Y_{ll} &=& C_G \left[ 
-S_{l-1} -S_{n-l+1} -{ 1 \over 2l} -{1 \over 2(n-l+2)}
+{ (l-1)(n+2) \over 2(n-2l+2)[n-l+1]_2} \right. \nonumber \\[8pt]
& & \left.  
-{ 2 \over n-l+2 } - { 1 \over l } \right. \nonumber \\[8pt]
& & \left. + { l-3 \over 2[l-1]_3} + \frac{l+2}{2[l-1]_2} 
           + { (n-l-1)(n-2l+4) \over 2[n-l+1]_3 }
+ { n-l+4 \over [n-l+1]_2 }
\right]
\nonumber \\[8pt]
& &+2(C_G-2C_F)\left[(-1)^l \left( {1 \over [l-1]_3 }+ 
{ (-1)^n +\ _{n-l+1}C_{l-1} \over [n-l+1]_3 } \right)\right.
\nonumber \\[8pt]
& &\left.-(-1)^{n} { l-1 \over 2(n-l+1)(n-2l+2) }\right]
\nonumber \\[8pt]
& &-C_F \left( 2S_{l-1} + 2S_{n-l+1} -3 \right)\ \ \ \ \
\left(   2 \le l \le \left\langle 
 {n \over 2} \right\rangle \right),
\label{eq3.13}
\end{eqnarray}
\begin{eqnarray}
Y_{{n \over 2}+1 \ {n \over 2}+1} &=& C_G \left[
      - 2S_{n \over 2} - {6 \over n+2}
      + { {n \over 2}-2 \over \left[{n \over 2}\right]_3 }
      + { {n \over 2}+3 \over \left[{n \over 2}\right]_2 } \right] 
      +(2C_F-C_G)  4 {(-1)^{n \over 2}  \over
                          \left[ {n \over 2} \right]_3 } 
\nonumber\\[8pt]
& & +C_F \left( 3 - 4S_{n \over 2} \right) \ \ \ \ 
\ \ \ \ \ \ \ ({\rm for\ even}\ n ),
\label{eq3.14}
\end{eqnarray}
\begin{eqnarray}
Y_{lm} &=& C_G \left[
  { (2n-l-m+4)(n-m+1) \over 2(m-l)[n-l+1]_2 }
+ { (m-1)(m+n-l+2) \over 2(n-m-l+2)[n-l+1]_2 }
\right. \nonumber\\[8pt]
& & \left. - { 2 \over n-l+2 }
+ { (n-l-1)(n-2l+4) \over 2[n-l+1]_3 }
+ { n-l+4 \over [n-l+1]_2 } 
\right] \nonumber\\[8pt]
& & + (C_G-2C_F) \left[{ 2 (-1)^{m} \left\{ _{n-l+1}C_{m-1} 
+ (-1)^n
\ _{n-l+1}C_{m-l}
\right\} \over
[n-l+1]_3 }\right.
\nonumber\\[8pt]
& & \left. -(-1)^{m-l} \left(
{ \ _{n-l}C_{m-l} \over (m-l)\ _{m-1}C_{m-l} }
+ { (-1)^{n}\ _{n-l}C_{m-2} 
\over (n-l+2-m)\ _{n-m+1}C_{l-1} }\right)
\right]
\nonumber\\[8pt] 
& &\left(
3 \le  l+1 \le m \le \left\langle {n \over 2}\right\rangle
\right),
\label{eq3.15}
\end{eqnarray}
\begin{eqnarray}
Y_{l \  {n \over 2}+1 } &=& C_G \left[
    {1 \over 4} {1 \over [n-l+1]_2} {n(3n-2l+6) \over n-2l+2}
  - {1 \over n-l+2} \right. \nonumber\\[8pt]
& & \left.  + {1 \over 2} \left( {(n-l-1)({n\over 2}-l+2) 
\over [n-l+1]_3 } 
                      +{n-l+4 \over [n-l+1]_2} \right) \right]
\nonumber\\[8pt]
& & +(2C_F-C_G) \left( {2(-1)^{n \over 2} \over [n-l+1]_3 } 
                    \ _{n-l+1}C_{n \over 2}
                    + { (-1)^{{n \over 2}-l+1} 
\over {{n \over 2}-l+1}}
{ {\ _{n-l}C_{{n \over 2}-l+1}} \over {\ _{n \over 2}C_{l-1}}}
 \right) \nonumber\\[8pt]
& &\ \ \ \ \ \left( {\rm for\ even}\ n,
\ \ 2 \le l \le  {n \over 2} \right),
\label{eq3.16}
\end{eqnarray}
\begin{eqnarray}
Y_{lE} =  -2C_F \left( {1 \over [l]_{2} } + { 1 \over [n-l+2]_2} 
 \right)\ \ \ \ \
\left( 2 \le l \le \left[{n\over 2}\right]+1 \right),
\label{eq3.17}
\end{eqnarray}
\begin{eqnarray}
Y_{lN}= 4C_F \left( {1 \over [l-1]_3 } + {1 \over [n-l+1]_3} 
\right)\ \ \ \ \
\left( 2 \le l \le \left[{n\over 2}\right]+1 \right),
\label{eq3.18}
\end{eqnarray}
\begin{eqnarray}
Y_{EE} = 2 (1-S_n)C_F,
\label{eq3.19}
\end{eqnarray}
\begin{eqnarray}
Y_{NN}=C_F\left({2 \over n(n+1) } -4S_n \right),
\label{eq3.20}
\end{eqnarray}
where $C_F={N_c^2-1\over 2N_c}$ and $C_G=N_c$ are the 
Casimir operators
of the gauge group $SU(N_c)$, 
$S_{n} = \sum_{j=1}^{n}1/j$,
$[j]_{k} = j(j+1)\cdots(j+k-1)$, and $_nC_l$ is the 
binomial coefficient
defined as $_nC_l=n!/l!(n-l)!$.
Note that (\ref{eq3.16}) is one half of (\ref{eq3.15}) with
$m={n\over 2}+1$ for even $n$. 
With these $Y_{ij}$ 
($i,j = 2,3,...,\left[ {n \over 2}\right]+1,\ E,\ N$),
the anomalous dimension matrix for the twist-3 operators
$R_{n,l}$, $E$ and $N$ take the form of 
the upper triangular matrix as
\begin{eqnarray}
\gamma_{ij} = - {g^2 \over 8\pi^2 } Y_{ij}.
\label{eq3.21}
\end{eqnarray}
The above results for $Y$ can be compared with the 
mixing matrix $X$
for $h_L$ given in Eqs. (3.14)-(3.20) of \cite{KT}.
Their difference comes from the opposite charge 
conjugation symmetry
of $R_{n,l}$. 
Solving the renormalization group equation,
the $Q^2$ evolution 
of $R_{n,l}$ and $N$ 
is given by
\begin{eqnarray}
a_{n,l}(Q^2) &=& 
\sum^{[\frac{n}{2}]+1}_{m=2}\left[ \left( 
{\alpha(Q^2) \over \alpha(\mu^2)}
\right)^{-Y /b_0} \right]_{lm}a_{n,m}(\mu^2)
+ \left[ \left( 
{\alpha(Q^2) \over \alpha(\mu^2)}
\right)^{-Y /b_0} \right]_{lN}d_{n}(\mu^2),
\label{eq3.24}\\
d_{n}(Q^{2}) &=& \left( 
{\alpha(Q^2) \over \alpha(\mu^2)}
\right)^{-Y_{NN} /b_0} d_{n}(\mu^2),
\label{eq3.25}
\end{eqnarray}
where 
$a_{n,l}$ and
$d_{n}$ are defined by
\begin{eqnarray}
\langle P | R_{n,l}^{\mu_1 \cdots \mu_n}(\mu^2) | P \rangle 
&=& 2a_{n,l}(\mu^2) M {\cal S}_n ( P^{\mu_1} 
\cdots P^{\mu_n} - {\rm traces}),
\label{eq3.22}\\
\langle P | N^{\mu_1 \cdots \mu_n}(\mu^2) | P \rangle 
&=& 2 d_{n}(\mu^2) M {\cal S}_n ( P^{\mu_1} 
\cdots P^{\mu_n} - {\rm traces}),
\label{eq3.23}
\end{eqnarray}
and $b_0={11\over 3}N_c -{2\over 3}N_f$.

The complete spectrum of the eigenvalues of the 
anomalous dimension matrix
(\ref{eq3.21}) (ignoring the factor $g^2/8\pi^2$)
 is shown in Fig. 4(a) 
together with those for the twist-2 distribution $h_1$.
One sees from the figure that the anomalous dimensions of $e$
are significantly larger than those
of $h_1$, except for the first several moments.

Before closing this section, we compare the $Q^2$ evolution among
the twist-2 and -3 distributions (at $m_q=0$), 
using the first few moments.
For $n=2$ and $3$, only one operator contributes to
$e(x)$.  The values of $Y$ can be read from (\ref{eq3.13}) and 
(\ref{eq3.14}) as $Y^{n=2}={-55\over 9}$ and $Y^{n=3}={-73\over 9}$.
Thus the $Q^2$ evolution becomes
\begin{eqnarray}
{\cal M}_2\left[ e(Q^2)\right]=L^{6.11/b_0}
{\cal M}_2\left[ e(\mu^2)\right],\qquad
{\cal M}_3\left[ e(Q^2)\right]=L^{8.11/b_0}
{\cal M}_3\left[ e(\mu^2)\right],
\end{eqnarray}
where $L={\alpha(Q^2)\over \alpha(\mu^2)}$.
In table 1, we summarize 
the anomalous dimensions of the twist-2 and -3
distributions for $n=2$, $3$, ignoring the common 
factor $g^2/8\pi^2$.
One sees that these moments of $e$ evolve slower than those
of the other twist-3 distributions 
and are close to the twist-2 distributions.
\begin{center}
\begin{tabular}{|c|c|c|c|c|c|} \hline
 n & $f_1$, $g_1$    & $\ h_1\ $ & $\ e\ $  
& $\ \tilde{g}_2\ $ & $\ \tilde{h}_L\ $ \\ \hline 
 & & & & & \\
 2 & ${50\over 9}$   & ${52\over 9}$ & ${55\over 9}$ & ${77\over 9}$ 
& $-$ \\ 
 & & & & & \\
 3 & ${314\over 45}$ & ${64\over 9}$ & ${73\over 9}$  & $-$
& ${104\over 9}$  \\ 
 & & & & & \\ \hline
\end{tabular}
\end{center}

\noindent
{\bf Table 1.} The anomalous dimensions for the 2nd and 3rd moments
of the nonsinglet 
twist-2 and twist-3 distributions.  
For these moments, 
the twist-3 distributions 
receive the contribution from only one operator.
$\tilde{h}_L$ and $\tilde{g}_2$
denote the twist-3 parts of $h_L$ and $g_2$, respectively.
For $\tilde{h}_L$, the 
third moment is the lowest one, and the result is
taken from \cite{KT}.  
For $g_2$, the result is taken from \cite{BKL}, and
the anomalous dimensions 
for odd moments are not available because they have
been discussed in the context of DIS.

\section{Large $N_c$ limit}
\setcounter{equation}{0}
\renewcommand{\theequation}{\arabic{section}.\arabic{equation}}

In \cite{ABH, BBKT}, it has been proved that
all the (nonsinglet) twist-3 distributions $g_2$, $h_L$ and $e$
obey a simple GLAP equation\,\cite{GLAP} similarly
to the twist-2 distributions in the $N_c\to\infty$ limit.
In this limit, $Q^2$ evolution of these distributions
is completely determined
by the lowest eigenvalue of the anomalous dimension
matrix which has a simple analytic form.  For $e$, 
this can be checked using the results in the previous section.
At $N_c\to\infty$, i.e., $C_F\to N_c/2$, 
the lowest
eigenvalue of $\gamma$ in (\ref{eq3.21}) is 
given by (ignoring the factor $g^2/8\pi^2$)
\begin{equation}
\gamma_n^e = 2N_c \left( S_n - \frac{1}{4} - \frac{1}{2(n+1)} \right).
\label{eq4.1}
\end{equation}
As was shown in section 2,
the $n$-th moment of $e(x)$ can be expressed in terms of
$a_{n,l}$ as
\begin{eqnarray}
{\cal M}_n\left[ e\right]
=\sum_{l=2}^{\langle{n\over 2}\rangle} a_{n,l}
\left\{ +{1\over 2}a_{n,\left[{n\over 2}\right]+1}\ \ \ 
n:\ {\rm even}
\right\},
\label{eq4.2}
\end{eqnarray}
where $\{ \cdots\ n:\ {\rm even}\}$ means this term is
only for even $n$.  
We can directly check that the coefficients
of $a_{n,l}$ in (\ref{eq4.2}) constitute the
{\it left} eigenvector of the mixing matrix $Y$ with 
the eigenvalue
$-\gamma_n^e$, i.e.,
\begin{eqnarray}
\sum_{l=2}^{\left[ {n\over 2}\right]+1}Y_{lm}=-\gamma_n^e
\qquad \left( m=2,\cdots,\left[ {n\over 2}\right] +1 \right)
\end{eqnarray}
for odd $n$ and 
\begin{eqnarray}
& &\sum_{l=2}^{\left[ {n\over 2}\right]}Y_{lm}
+{1\over 2}Y_{{n\over 2}+1\ m}
=-\gamma_n^e
\qquad \left( m=2,\cdots,\left[ {n\over 2}\right] \right),
\nonumber\\
& &\sum_{l=2}^{\left[ {n\over 2}\right]}Y_{l\ {n\over 2}+1}
+{1\over 2}Y_{{n\over 2}+1\ {n\over 2}+1}
=-{1\over 2}\gamma_n^e
\end{eqnarray}
for even $n$.  This
means all the {\it right} eigenvectors except the one 
which corresponds to
$\gamma_n^e$ are orthogonal to the above vector consisting of
the coefficients of $a_{n,l}$ in (\ref{eq4.2}).
This proves that
the $Q^2$ evolution of $e$ is exactly given by
\begin{eqnarray}
{\cal M}_n\left[ e(Q^2)\right]=L^{\gamma_n^e/b_0}
{\cal M}_n\left[ e(\mu^2)\right],
\end{eqnarray}
as is found in \cite{BBKT} by a different method.
The complete spectrum of the eigenvalues of the anomalous 
dimension matrix
at large $N_c$ 
is given in Fig. 4(b) 
together with the analytic result in (\ref{eq4.1}).
One sees that (\ref{eq4.1}) indeed corresponds 
to the lowest eigenvalue.
 
To see how the approximation by the $N_c\to\infty$ result works,
we compare the two results 
for $n=4$, as an example.
The exact mixing matrix is
\begin{eqnarray} 
Y =\left(\matrix{ -{511 \over 45}& {21 \over 10}\cr 
{10 \over 3}& 
-{122 \over 9}\cr}\right),
\label{eq4.3p}
\end{eqnarray} 
and the one at $N_c\to\infty$ is
\begin{eqnarray} Y =
\left(\matrix{ -{119 \over 10}& {9 \over 5}\cr 
{3}& 
-{14}\cr}\right).
\label{eq4.3}
\end{eqnarray} 
From these $Y$s' one gets 
\begin{eqnarray} 
{\cal M}_4\left[ e(Q^2)\right]
&=& \left(
0.983 a_{4,2}(\mu) + 0.520a_{4,3}(\mu) \right) 
\left({ \alpha(Q^2) \over
\alpha(\mu^2) }\right)^{9.59/b_0} \nonumber\\ &+& 
\left( 0.017 a_{4,2}(\mu) - 0.020a_{4,3}(\mu) \right) 
\left({ \alpha(Q^2) 
\over \alpha(\mu^2)
}\right)^{15.3/b_0}
\label{exact}
\end{eqnarray} 
for the exact $Q^2$ evolution and 
\begin{eqnarray} 
{\cal M}_4\left[ e(Q^2)\right] = \left(
a_{4,2}(\mu) + \frac{1}{2}a_{4,3}(\mu) \right) 
\left({ \alpha(Q^2) \over
\alpha(\mu^2) }\right)^{10.4/b_0}
\label{largenc}
\end{eqnarray} 
at $N_c\to\infty$.
Equations (\ref{exact}) and (\ref{largenc}) can be compared with
Eqs. (20) and (21) of \cite{BBKT} for $h_L$.
One sees here again that the coefficients in the 
second term of (\ref{exact})
are small (i.e., $1/N_c^2$ suppressed) and the 
anomalous dimension in
(\ref{largenc}) is close to the smaller one in (\ref{exact}).

\section{Summary}
\setcounter{equation}{0}
\renewcommand{\theequation}{\arabic{section}.\arabic{equation}}

In this paper, we presented the covariant calculation of
the anomalous dimension matrix
for the chiral-odd twist-3 distribution $e(x,Q^2)$.
The operator mixing with the EOM operator is taken into account.
This study completes the whole list of the anomalous dimensions
of all the twist-3 distributions in the one-loop level
together with the known results for $g_2$ and $h_L$. 
At large $N_c$, we have confirmed that the lowest eigenvalues of
the anomalous dimension matrix take a simple analytic form shown in 
(\ref{eq4.1}) and it governs the whole $Q^2$ evolution of $e(x,Q^2)$
as was found in \cite{BBKT} by a different method.

\vspace{1.0cm}
\noindent
{\large\bf Acknowledgement}

\vspace{0.5cm}

We thank V.M. Braun and K. Tanaka for useful discussions.

\vspace{1.0cm}
\appendix

\noindent
{\large\bf Appendix}

\setcounter{section}{1}
\setcounter{equation}{0}
\renewcommand{\theequation}{\Alph{section}.\arabic{equation}}

\vspace{0.5cm}

In this appendix we present the contribution from each one-loop
Feynman diagram shown in Figs. 2 and 3.

First we consider the three-point function $F_\mu\Omega^\mu$  with 
the insertion of 
$R_{n,l}\cdot \Delta$.  We set $m_q=0$ to get $Z_{lm}$ and $Z_{lE}$
($l,m=2,\ldots [n/2]+1$).

\noindent
Fig.2(a) gives 0.

\noindent
Fig.2(b) gives
\noindent
\begin{eqnarray}
\lefteqn{\frac{g^{2}}{16\pi^{2}\varepsilon}
C_{G} \left[ \sum_{m = 2}^{l-1} \left(\frac{(l+m)(m-1)}
{2 (l-m)[l-1]_{2}}
+ \frac{(n-l+2+m)(m-1)}{2(n-l+2-m)[n-l+1]_{2}} \right)
{\cal R}^{(3)}_{n,m} \right.} \nonumber \\
& & + \left( 1 - S_{l-1} - S_{n-l+1} 
- \frac{1}{2 l} - \frac{1}{2(n-l+2)} 
+ \frac{(n+2)(l-1)}{2(n-2l+2)[n-l+1]_{2}}\right) 
{\cal R}^{(3)}_{n,l}
\nonumber \\
& &+ \left. \sum_{m = l+1}^{\langle n/2 \rangle}
\left(
\frac{(2n-l-m+4)(n-m+1)}{2(m-l)[n-l+1]_{2}}
+ \frac{(n-l+2+m)(m-1)}{2(n-l+2-m)[n-l+1]_{2}}\right)
{\cal R}^{(3)}_{n,m} \right. \nonumber \\
& & \left. \left\{ +
\frac{1}{4} \frac{1}{[n-l+1]_2} \frac{n(3n-2l+6)}{n-2l+2}
 {\cal R}^{(3)}_{n,\frac{n}{2}+1}\ \ \ 
 n:\ {\rm even} 
\right\}
\right]\ \ \ \ 
\left( 2 \le l \le \left\langle \frac{n}{2} \right\rangle \right),
\label{eqb}
\end{eqnarray}
where $\{\cdots\  n:\ {\rm even} \}$ means this term
is only for an even $n$ 
and its coefficient is one half of that of the third 
line with $m=[n/2]+1$ 
(the same in
(\ref{eqcd}), (\ref{eqef}) and (\ref{eqgh}) below), and
\begin{eqnarray}
\lefteqn{\frac{g^{2}}{16\pi^{2}\varepsilon}C_{G} 
\left[ 
\sum_{m=2}^{[{n\over 2}]}\left( 
{2\over {n\over 2}-m+1} - { n+m+1 \over \left[ {n\over 2}\right]_2 }
\right)
{\cal R}^{(3)}_{n,m}
\right.}
\nonumber \\
& & \left. + \left( 1 - S_{n\over 2} - S_{{n\over 2}+1}
\right) 
{\cal R}_{n, \frac{n}{2}+1}^{(3)} \right]\ \ \ \
\left( {\rm for}\ l=[\frac{n}{2}] +1 \ 
{\rm with\ even}\ n\right).
\label{eqb2}
\end{eqnarray}
In (\ref{eqb2}), the terms with $m=2,\ldots, [n/2]$ are  
the same as the first line of (\ref{eqb}) with $l=[n/2]+1$.
This rule also applies to the results for other diagrams.

\noindent
Fig.2(c) + Fig.2(d) gives
\begin{eqnarray}
\lefteqn{\frac{g^{2}}{16\pi^{2}\varepsilon}(2 C_{F}-C_{G})
\left[
\sum_{m=2}^{l-1}2(-1)^{m-1}\left( \frac{_{n-l+1}C_{m-1}}
{[n-l+1]_{3}}
+ \frac{_{l-1}C_{m-1}}{[l-1]_{3}}\right) 
{\cal R}_{n,m}^{(3)}\right.} 
\nonumber \\
& &+2(-1)^{l-1} \left( \frac{_{n-l+1}
C_{l-1} + (-1)^{n}}{[n-l+1]_{3}}
+ \frac{1}{[l-1]_{3}} \right) {\cal R}_{n,l}^{(3)}  
\nonumber\\
& &+ \sum_{m = l+1}^{\langle n/2\rangle}
\frac{2(-1)^{m-1}}{[n-l+1]_{3}}
\left(_{n-l+1}C_{m-1} + (-1)^{n}\: _{n-l+1}C_{m-l}\right) 
{\cal R}^{(3)}_{n,m} \nonumber \\
& &  \left\{ +
 \frac{2(-1)^{\frac{n}{2}}}{[n-l+1]_3} \ _{n-l+1}C_{\frac{n}{2}}
 {\cal R}^{(3)}_{n,{\frac{n}{2}+1}}\ \ \ \  n:\ {\rm even} \right\}
\nonumber \\
& & -\left. \left(\frac{1}{[l]_{2}} + 
\frac{1}{[n-l+2]_{2}}\right) 
{\cal E}^{(3)}_{n} \right]\ \ \ \ \ 
\left( 2 \le l \le \left\langle \frac{n}{2} \right\rangle \right),
 \label{eqcd}
\end{eqnarray}
and
\begin{eqnarray}
& &\frac{g^{2}}{16\pi^{2}\varepsilon}
(2C_F-C_G)\left[ \sum_{m=2}^{[n/2]+1}4(-1)^{m-1} 
\frac{_{n\over 2}C_{m-1}}
{[n/2]_{3}}
{\cal R}_{n,m}^{(3)}  
 - { 2 \over \left[ {n\over 2} +1\right]_2}
{\cal E}_{n}^{(3)} \right]\nonumber\\
& &\qquad\qquad\qquad \left( {\rm for}\ 
l=\left[ \frac{n}{2} \right] +1
\ {\rm with\ even}\ n \right). 
\end{eqnarray}

\noindent
Fig.2(e) + Fig.2(f) gives
\begin{eqnarray}
\lefteqn{\frac{g^{2}}{16\pi^{2}\varepsilon}\left[
\sum_{m = 2}^{l-1}\left\{ (2C_{F}-C_{G})
\left(\frac{(-1)^{l-m}\: 
_{l-2}C_{l-m}}{(l-m)\: _{n-m+1}C_{l-m}} +
\frac{(-1)^{n-l-m}\: _{n-l}C_{m-2}}{(n-l+2-m)
\: _{n-m+1}C_{l-1}}\right)
\right. \right.} \nonumber \\
& &- \left. C_{G}\left( \frac{1}{l} + \frac{1}{n-l+2}
\right)\right\} 
{\cal R}_{n,m}^{(3)} 
\nonumber \\
& &+ \left\{ (2C_{F}-C_{G}) \frac{(-1)^{n}(l-1)}
{(n-2l+2)(n-l+1)}
-2C_{F}(S_{l-1} + S_{n-l+1} - 2) \right. \nonumber \\
& &- C_{G} \left. \left( \frac{1}{l} + \frac{2}{n-l+2} \right)
\right\} {\cal R}_{n,l}^{(3)} \nonumber \\
& &+ \left\{ \sum_{m = l+1}^{\langle n/2\rangle} (2C_{F}- C_{G}) \left(
\frac{(-1)^{m-l}\: _{n-l}C_{m-l}}{(m-l)\: _{m-1}C_{m-l}}
+ \frac{(-1)^{n-l-m}\: _{n-l}C_{m-2}}
{(n-l+2-m) _{n-m+1}C_{l-1}}
\right) \right. \nonumber \\ 
& &- 
\left. C_G\frac{2}{n-l+2} \right\} {\cal R}_{n,m}^{(3)} \nonumber \\
& &  \left\{ +
\left\{ (2C_F-C_G) \left( 
\frac{(-1)^{\frac{n}{2}-l+1}}{\frac{n}{2}-l+1}
  \frac{ _{n-l}C_{\frac{n}{2}-l+1}}
{ _\frac{n}{2}C_{\frac{n}{2}-l+1}} \right)
- C_G\frac{1}{n-l+2} \right\} {\cal R}_{n,\frac{n}{2}+1}^{(3)}\ \ \ \ 
n:\ {\rm even} \right\}
 \nonumber \\
& & - \left.  C_{G}\left(\frac{1}{l} + \frac{1}{n-l+2}
\right){\cal E}_{n}^{(3)} \right]\ \ \ \ 
\qquad  \left( 2 \le l \le \left\langle \frac{n}{2} \right\rangle
\right),
\label{eqef}
\end{eqnarray}
and
\begin{eqnarray}
\lefteqn{\frac{g^{2}}{16\pi^{2}\varepsilon}\left[
\sum_{m = 2}^{n/2}\left\{ (2C_{F}-C_{G})
\frac{2(-1)^{{n\over 2}+1-m}\: 
_{{n\over 2}-1}C_{m-2}}{({n\over 2}+1-m)\: _{n-m+1}C_{n \over 2}}
- C_{G} \frac{2}{{n\over 2}+1} \right\} 
{\cal R}_{n,m}^{(3)} 
\right. } \nonumber \\
& & + \left. \left\{ 
4C_F(1-S_{\frac{n}{2}}) - C_G \frac{2}{\frac{n}{2}+1}
 \right\} {\cal R}_{n  \frac{n}{2}+1 }^{(3)} 
 - C_G \frac{2}{ {n\over 2}+1} {\cal E}_n^{(3)} 
\right] \nonumber\\
& &\qquad\qquad\qquad\left({\rm for}\ l=\frac{n}{2}+1\ 
{\rm with\ even}\ n \right).
\end{eqnarray}

\noindent
Fig.2(g) + Fig.2(h) gives
\begin{eqnarray}
\lefteqn{\frac{g^{2}}{16\pi^{2}\varepsilon}
C_{G}\left[\sum_{m=2}^{l-1} \frac{1}{2}\left(
\frac{(l-3)(l+1-m)}{[l-1]_{3}} + \frac{l+2}{[l-1]_{2}}
\right.\right.} \nonumber \\
& &+\left.\frac{(n-l-1)(n-l+3-m)}{[n-l+1]_{3}} 
+ \frac{n-l+4}{[n-l+1]_2}
\right) {\cal R}_{n,m}^{(3)} 
\nonumber \\
& & + \frac{1}{2}\left( \frac{l-3}{[l-1]_{3}} 
+ \frac{l+2}{[l-1]_{2}}
+ \frac{(n-l-1)(n-2l+4)}{[n-l+1]_3}
+ 2 \frac{n-l+4}{[n-l+1]_2} \right) {\cal R}_{n,l}^{(3)} 
\nonumber \\
& & + \left.\sum_{m=l+1}^{\langle n/2\rangle} 
\frac{1}{2} \left( \frac{(n-l-1)(n-2l+4)}{[n-l+1]_3}
    + \frac{2(n-l+4)}{[n-l+1]_2} \right) {\cal R}_{n,m}^{(3)} \right.
 \nonumber \\
& &  \left\{ +
\frac{1}{2} \left( \frac{(n-l-1)({n\over 2}-l+2)}{[n-l+1]_3}
+ \frac{n-l+4}{[n-l+1]_2} \right) {\cal R}_{n,\frac{n}{2}+1}^{(3)}
\ \ \ \ n:\ {\rm even}
\right\} 
\nonumber \\
& & + \left.\left(\frac{1}{l+1} + \frac{1}{n-l+3} 
\right) {\cal E}^{(3)}_{n}
\right] \ \ \ \ \ \ \left( 2 \le l \le \left\langle 
\frac{n}{2}\right\rangle \right),
\label{eqgh}
\end{eqnarray}
and
\begin{eqnarray}
& &\frac{g^{2}}{16\pi^{2}\varepsilon}
C_{G}\left[\sum_{m=2}^{{n\over 2}+1}
\left( \frac{ \left(\frac{n}{2}-2\right)
\left( {n\over 2} +2 -m   \right)     }{[\frac{n}{2}]_3}
+ \frac{\frac{n}{2}+3}{[\frac{n}{2}]_2} \right) 
{\cal R}_{n,m}^{(3)}
+  \frac{2}{{n\over 2}+2} {\cal E}_{n}^{(3)}
\right]\nonumber\\
& & \qquad\qquad\qquad \left({\rm for}\ l=\frac{n}{2}+1
\ {\rm with\ even}\ n \right). 
\end{eqnarray}

Next we calculate $Z_{lE}$, $Z_{lN}$, $Z_{EE}$ and $Z_{NN}$ from
the one-loop two-point functions shown in Fig. 3
with a nonzero quark mass.

The one-loop correction to the two-point function with 
$R_{n,l}\cdot \Delta$ comes 
from Fig. 3 (b) and (c). It gives
\begin{eqnarray}
\lefteqn{\frac{g^{2}}{16 \pi^{2} \varepsilon}C_{F}
\left[ - \left(\frac{2}{[l]_{2}}+
\frac{2}{[n-l+2]_{2}}\right)
{\cal E}_n^{(2)}\right.}
\nonumber\\
& &+ \left.\left(\frac{4}{[l-1]_{3}}+\frac{4}
{[n-l+1]_{3}}\right)
{\cal N}_n^{(2)} \right].
\label{eq:a10}
\end{eqnarray}

For the one-loop correction to the two-point function with 
$E_{n}\cdot\Delta$,
Fig.3(a) gives
\begin{equation}
\frac{g^{2}}{16\pi^{2}\varepsilon}C_{F}
\left[ \frac{4}{n+1} {\cal E}_n^{(2)} +
2 \frac{n-1}{[n]_2} {\cal N}_n^{(2)} \right]
\label{eq:a11}
\end{equation}
and Fig.3(b)+Fig.3(c) gives
\begin{equation}
\frac{g^{2}}{16\pi^{2}\varepsilon}C_{F}
\left[ \left(3 - \frac{4}{n+1} - 2S_n
\right){\cal E}_n^{(2)}
- \left(3 + \frac{2(n-1)}{[n]_2}\right){\cal N}_n^{(2)}\right].
\label{eq:a12}
\end{equation}
Although each contribution from Fig. 3 (a) 
and (b) is different from
the one for $h_L$, the sum of
(\ref{eq:a11}) 
and (\ref{eq:a12}) are the same as the one for $h_L$.

Finally, for the one-loop correction to 
the two-point function with $N_{n}\cdot\Delta$,
Fig.3(a) gives
\begin{equation}
\frac{g^{2}}{16\pi^{2}\varepsilon} C_{F} \frac{2}{n(n+1)}
{\cal N}_n^{(2)},
\label{eq:a13}
\end{equation}
and each of Figs.3 (b) and (c) gives 
the same contribution,
\begin{equation}
\frac{g^{2}}{16\pi^{2}\varepsilon}C_{F}\left(- 2 
\sum_{j=2}^{n}
\frac{1}{j}\right) {\cal N}_n^{(2)}.
\label{eq:a14}
\end{equation}
These are the same as the ones for $h_L$.

\newpage
\centerline{\bf Figure captions}

\vspace{1cm}

\begin{enumerate}

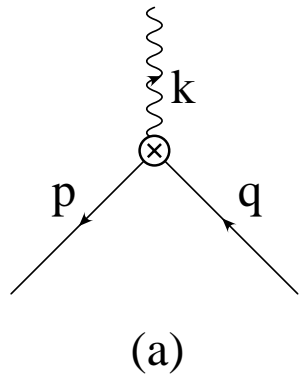
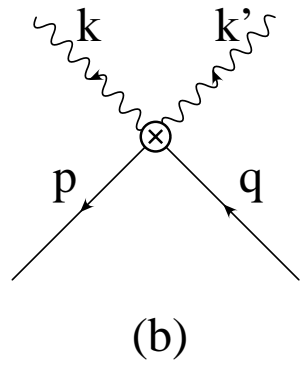
\item[{\bf Fig. 1.}] 
(a) Three-point basic vertex for $R_{n,l}$, $E$, and $N$.
(b) Four-point basic vertex for $R_{n,l}$ necessary for the 
calculation of the diagrams shown in Fig. 2.

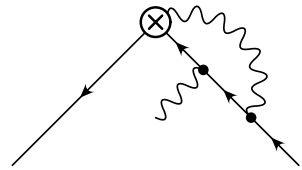
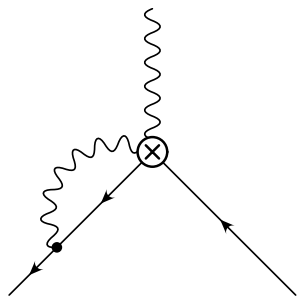
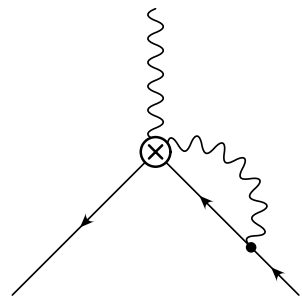
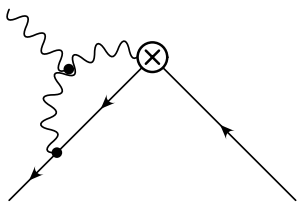
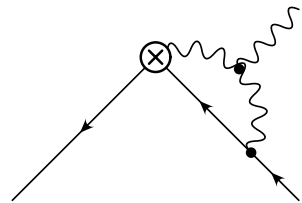
\item[{\bf Fig. 2.}] One-particle-irreducible diagrams for the 
one-loop correction to $F_\mu(p,q,k)$.

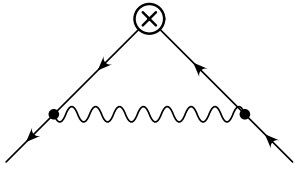
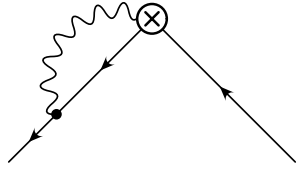
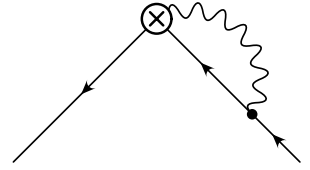
\item[{\bf Fig. 3.}] One-loop corrections to the 
two-point functions.

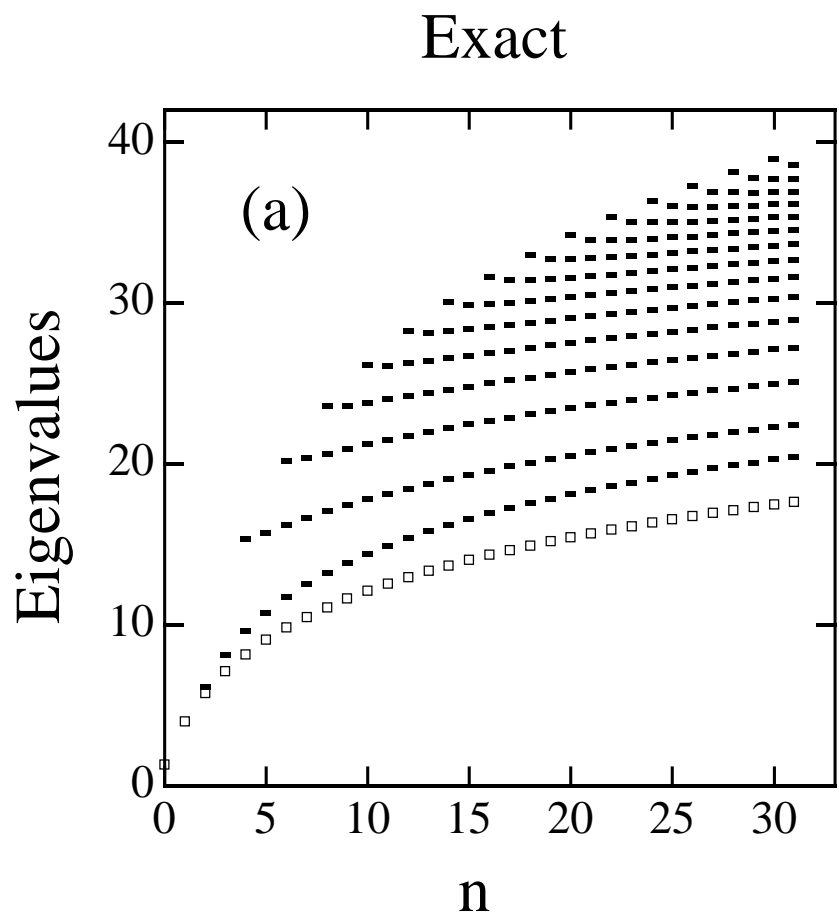
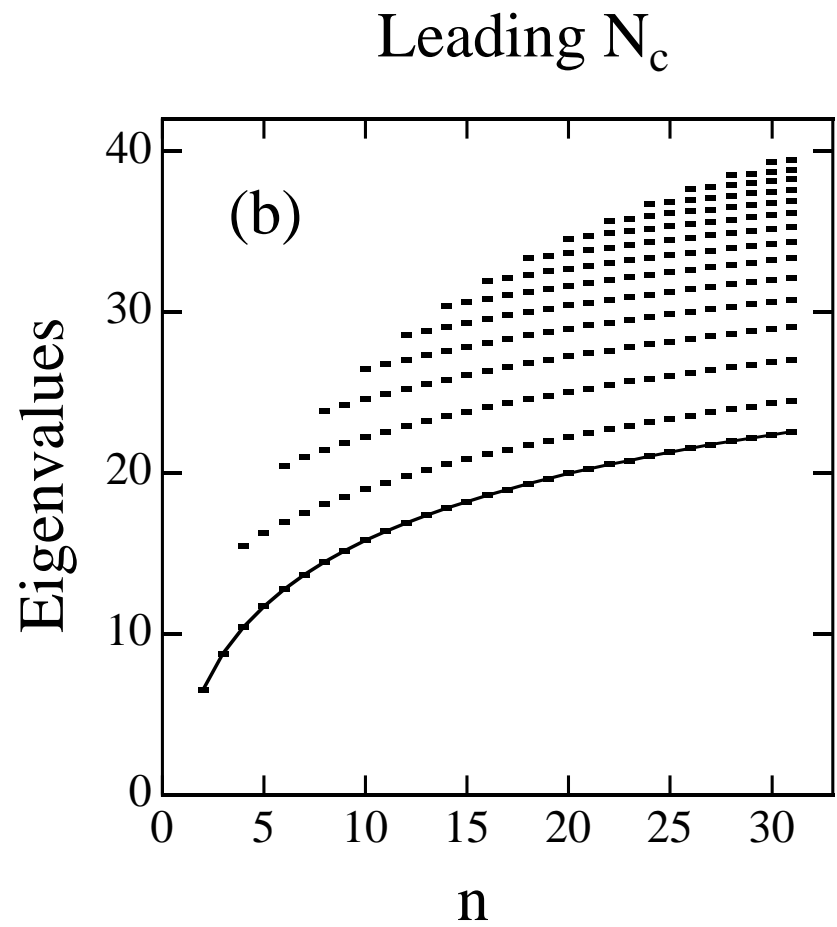
\item[{\bf Fig. 4.}] (a) Complete spectrum of the 
exact eigenvalues of the
anomalous dimension matrix for $e(x,Q^2)$ together 
with those for $h_1$
(squares).
(b) Complete spectrum of the leading $N_c$ eigenvalues of the
anomalous dimension matrix for 
$e(x,Q^2)$.
Solid line is the analytic
solution in (\ref{eq4.1}).

\end{enumerate}
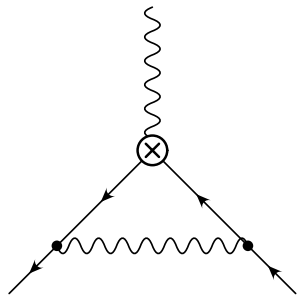
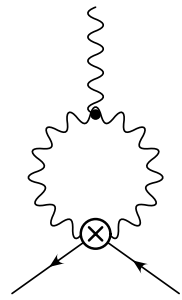
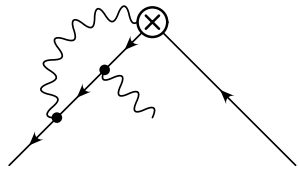
\end{document}